\documentclass{ws-jnopm}
\usepackage{graphicx}
\begin{document}

\markboth{T. Singh \& M. R. Shenoy}{Modeling and design consideriation of OPOs}

\catchline{}{}{}{}{}

\title{Modeling and design of singly-resonant optical parametric oscillator with an intracavity idler absorber for enhanced conversion efficiency for the signal}

\author{TAJINDER SINGH\footnote{Corresponding author.}} 

\address{Department of Nuclear and Atomic Physics, Tata Institute of Fundamental Research,\\
Mumbai, Maharashtra 400005, India.\\
tajinder.singh@tifr.res.in}

\author{M. R. SHENOY}

\address{Department of Physics, Indian Institute of Technology,\\
Hauz Khas, New Delhi 110016, India\\
mrshenoy@physics.iitd.ac.in}

\maketitle

\begin{history}
\end{history}

\begin{abstract}
We present modeling and design of singly-resonant optical parametric oscillator (SR-OPO) with an intracavity idler absorber to enhance the conversion efficiency for the signal, by suppressing the back conversion of the signal and idler to the pump. Following plane wave analysis, we arrive at the optimum parameters of the OPO to achieve high conversion efficiency for the signal. For a given pump intensity, we have analyzed the effect of position and number of absorbers required for optimum performance of the device. The model is also extended to the case in which the signal is absorbed, yielding higher conversion efficiency for the idler (in mid-IR region). The magnitude of absorption and the effect of inter-crystal phase shift on the conversion efficiency are also discussed. We also present an analytical solution for twin-crystal SR-OPO with an absorber in between; taking into account the variation of signal amplitude inside the cavity, we re-affirm that the often used ‘constant signal-wave approximation’ is valid if the reflectivity of the output coupler is high for the signal.
\end{abstract}


\section{Introduction}	

Optical parametric oscillators (OPOs) are widely developed and used as sources of tunable light in the visible and near IR region. There is a considerable amount of literature available on the modeling, analysis and experimentation of OPOs. Most of the existing models of OPOs dealt with their performance for pump power nearly five to six times the threshold. Beyond this pump power, the conversion efficiency was limited by back conversion (annihilation of signal and idler photons back to pump photons).$^{1-6}$ If the idler wave is absorbed inside the nonlinear crystal, then it would prevent the back conversion$^7$; i.e., due to the absence of idler wave, pump cannot build up by depleting the signal wave. But the nonlinear crystals may not possess large absorption coefficient at the idler wavelength. One of the ways to increase the absorption of idler wave is to dope the nonlinear crystal with absorbing impurities. However, increasing the idler absorption coefficient by doping could affect the nonlinear properties of the crystal. Also, the absorption of idler wave inside the nonlinear crystal causes the thermal loading$^{8,9}$ of the nonlinear crystal and results in a decrease in the conversion efficiency.$^{10-12}$ There are some other OPO configurations to suppress the back conversion, for example, using a two-crystal OPO with four dichroic mirrors.$^{13}$ But such configurations have some limitations, such as longer build-up time due to the use of ring cavity and more signal loss due to the use of four mirrors.$^{14}$

In this paper, we present a plane wave analysis of the twin crystal OPO, which employs a separate (stand-alone) absorber in between the two nonlinear crystals wherein the absorption of idler wave takes place. Use of a separate absorber has some additional benefits over the configuration in which absorption takes place inside the nonlinear crystal. By varying the position of the absorber in our model, we show that the back conversion is completely suppressed if it is placed nearly mid-way between the two crystals. Putting an absorber between the crystals will introduce an additional relative inter-crystal phase shift that can be used to cancel the effect of phase mismatch in the two crystals just like in parametric amplifiers.$^{15,16}$ A brief discussion of three-crystal configuration with two idler absorbers is also presented. Further, instead of using an idler absorber, if we use a signal absorber, then we can enhance the idler conversion efficiency. To the best of our knowledge, there are very few schemes that give high conversion efficiency in mid-IR region. For single crystal configuration, constant signal-wave approximation is valid when the reflectivity of the output mirror is high.$^{17}$ We have presented a similar analysis for the twin-crystal configuration, now with an idler absorber in between the two crystals.

\section{SR-OPO with uniform idler absorption}
Under the slowly varying envelope approximation, the three coupled differential equations,$^{18}$ for optical parametric generation including the idler absorption, are given below:
\begin{equation}
\dfrac{dE_\text{p}}{dz}=-i\kappa_\text{p}E_\text{i}E_\text{s}\exp{(i\Delta kz)}
,\label{this}
\end{equation}
\begin{equation}
\dfrac{dE_\text{s}}{dz}=-i\kappa_\text{s}E_\text{p}E^*_\text{i}\exp{(-i\Delta kz)}
,\label{this}
\end{equation}
\begin{equation}
\dfrac{dE_\text{i}}{dz}=-i\kappa_\text{i}E_\text{p}E^*_\text{s}\exp{(-i\Delta kz)}-\frac{\alpha_\text{i}}{2}E_\text{i}
,\label{this}
\end{equation}
where $E_\text{m}$'s are the amplitudes of the three waves (m=p, s, i). $\alpha_\text{i}$ is the idler absorption coefficient and
\begin{equation}
\kappa_\text{m}=\frac{\omega_\text{m}d_{\text{eff}}}{n_\text{m}c}
\label{this}
\end{equation}
are the coupling constants for the three waves; $d_{\text{eff}}$ is the effective nonlinear coefficient of the crystal, $\omega_\text{m}$'s are the angular frequencies and $n_\text{m}$'s are the corresponding refracive indices of the interacting waves and $c$ is the speed of light in vacuum. The phase mismatch term between the pump and the generated waves is given by:
\begin{equation}
\Delta k=k_\text{p}-(k_\text{s}+k_\text{i})
.\label{this}
\end{equation}
$k_\text{m}$'s are the magnitudes of the propagation vectors of the three waves, and the energy conservation requires:
\begin{equation}
\omega_\text{p}=\omega_\text{s}+\omega_\text{i}
.\label{this}
\end{equation}
Schematic of an SR-OPO with idler absorption inside the crystal is shown in Fig. 1. M$_1$ and M$_2$ are dichroic mirrors with high reflectivity for the signal and high transmitivity for the idler and pump. $L$ is the length of the crystal.
\begin{figure}[th]
\centering
\includegraphics[scale=0.4]{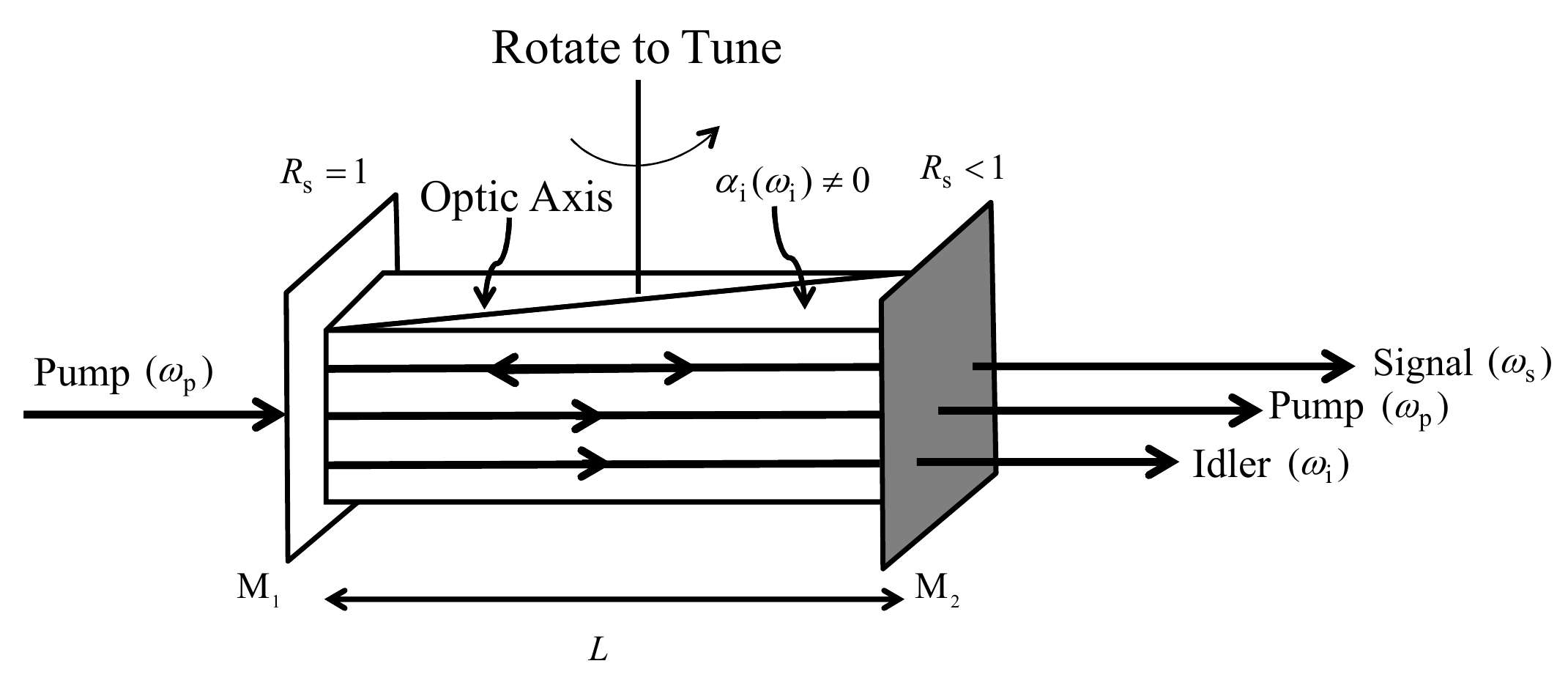}
\vspace*{8pt}
\caption{Schematic of SR-OPO with idler absorption inside the crystal.}
\end{figure}
For the case of perfect phase matching $(\Delta k=0)$ and constant signal wave intensity in the cavity (which is true if the reflectivity of the output coupler is large,$^{17}$ ($R_\text{s}>0.7$)), if we solve Eqs. (2.1)-(2.3) by standard procedure with the boundary conditions:
\begin{equation}
E_\text{p}(z=0)=E_\text{po} \quad \text{and} \quad E_\text{i}(z=0)=0
,\label{this}
\end{equation}
then we will get expressions for pump and idler amplitudes at any $z$ inside the crystal.

Since the signal is assumed to be in steady state, the number of signal photons lost per second in one round trip will be equal to the number of pump photons reduced per second in single pass through the cavity,$^{19}$ i.e.,
\begin{equation}
\frac{P_\text{s}}{\hbar \omega_\text{s}}\times \text{Signal loss}=\frac{P_\text{p}(0)-P_\text{p}(L)}{\hbar \omega_\text{p}}
,\label{this}
\end{equation}
where $\hbar=h/2\pi$, $h$ is the Planck's constant, $P_\text{m}$ (m=p, s, i) is the power of corresponding waves and the signal loss is defined as$^{19}$:
\begin{equation}
\text{Signal loss} = 1-\exp{\left(-{\frac{2 \Gamma_\text{s} n_\text{s} L}{c}}\right)}
.\label{this}
\end{equation}
$\Gamma_\text{s}$ is the damping constant for the signal wave, and is given by:
\begin{equation}
\Gamma_\text{s}=\frac{c}{n_\text{s}}\left( \alpha_\text{s}-\frac{1}{2L} \ln(R_\text{s}) \right)
,\label{this}
\end{equation}
where $\alpha_\text{s}$ is the absorption coefficient for the signal wave inside the crystal. In the practical cases $\alpha_\text{s}=0$ , and therefore neglecting $\alpha_\text{s}$ , Eqs. (2.7)-(2.10) gives:
\begin{equation}
\frac{1}{(\beta L)^2}\left(1-\left|  \frac{E^{\text{out}}_\text{p}}{E_\text{po}}\right|^2  \right)=\frac{1}{N_\text{p}}
.\label{this}
\end{equation}
The conversion efficiency for the signal is given by:
\begin{equation}
\eta_\text{s}=(1-R_\text{s})\frac{P_\text{s}}{P_\text{po}} \left( 1-\left|  \frac{E^{\text{out}}_\text{p}}{E_\text{po}}\right|^2  \right)
,\label{this}
\end{equation}
where
\begin{equation}
\beta^2=\kappa_\text{p}\kappa_\text{i}|E_\text{s}|^2
.\label{this}
\end{equation}
$\beta$ is the parametric gain of the OPO, $N_\text{p}$ is the input pump intensity normalized to its threshold value for $\alpha_\text{ai}=0$ , which is given by:
\begin{equation}
I^{\text{th}}_\text{p}(\alpha_\text{i}=0)=\frac{\lambda_\text{i} \lambda_\text{s} n_\text{p} n_\text{s} n_\text{i} c \epsilon_\text{o} (1-R_\text{s})}{8\pi^2d^2_{\text{eff}}L^2}
.\label{this}
\end{equation}
The threshold pump intensity is determined from the fact that near threshold $\beta L \rightarrow 0$.
\begin{equation}
N^{\text{th}}_\text{p}=\frac{I^{\text{th}}_\text{p}(\alpha_\text{i}\neq0)}{I^{\text{th}}_\text{p}(\alpha_\text{i}=0)}=\left( \frac{4}{\alpha_\text{i}L} -\frac{8}{\alpha^2_\text{i}L^2} \left( 1-\exp \left(  -\frac{\alpha_\text{i}L}{2} \right)  \right) \right)^{-1}
,\label{this}
\end{equation}
where $N^{\text{th}}_\text{p}$ is the normalized threshold pump intensity.
\begin{figure}[th]
\centering
\includegraphics[scale=0.4]{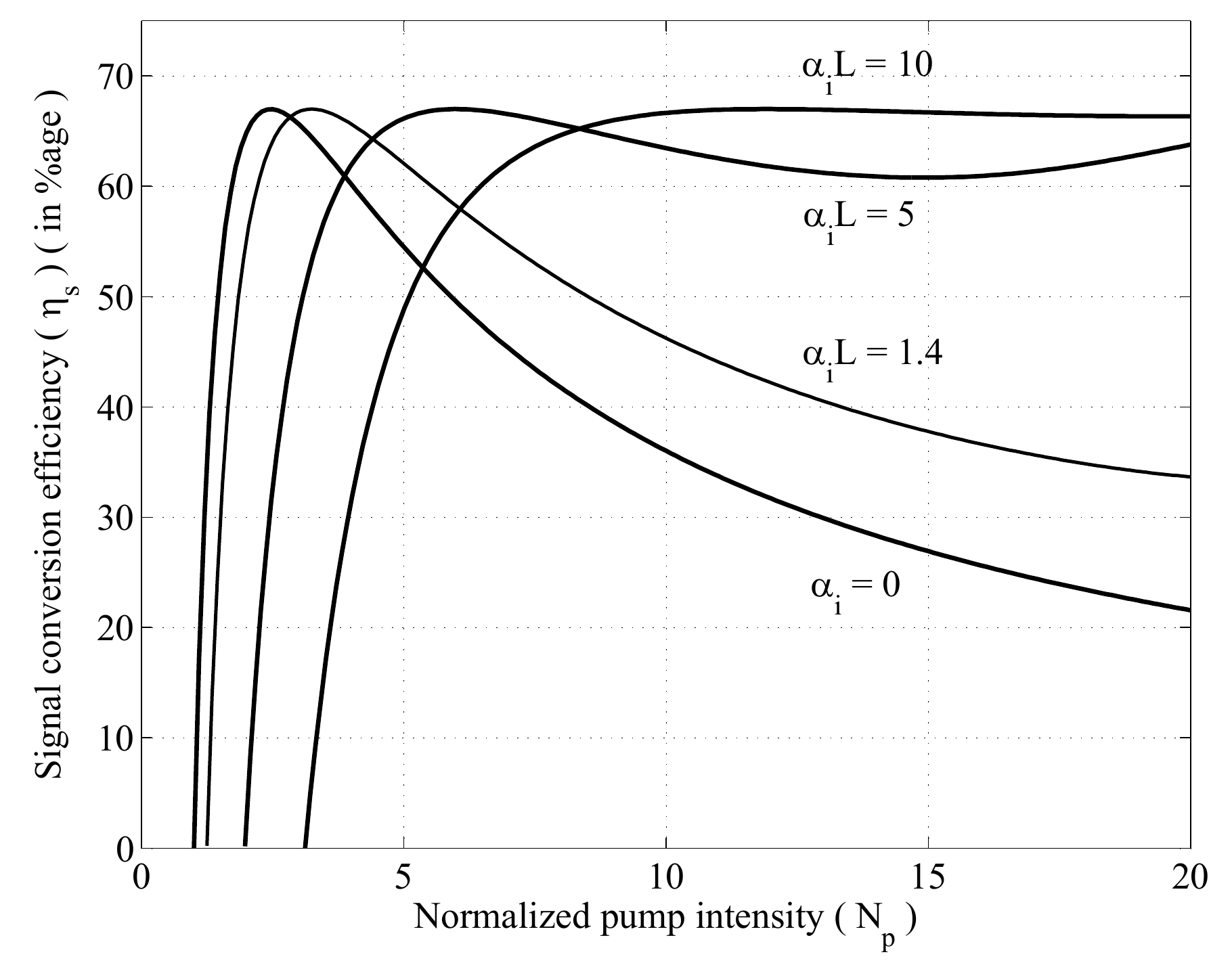}
\vspace*{8pt}
\caption{Variation of conversion efficiency for the signal with normalized pump
intensity for different values of $\alpha_\text{i}L$, $\alpha_\text{i}$ is the idler absorption coefficient in the nonlinear crystal.}
\end{figure}

Following a slightly different approach from Lowenthal,$^7$ we have obtained the variation of conversion efficiency for the signal with normalized pump intensity for different values of $\alpha_\text{i}L$ , which is shown in Fig. 2. The numerical results are presented for the wavelengths $\lambda_\text{p}=1.064\mu m$, $\lambda_\text{s}=1.59\mu m$, and $\lambda_\text{i}=3.216\mu m$. When $\alpha_\text{i}L$ is small, then the idler wave is slightly absorbed and keeps contributing to back conversion at high pump intensities. As $\alpha_\text{i}L$ increases, idler wave gets almost completely absorbed and there is no back conversion. Our results with $\alpha_\text{i}=0$ and $\alpha_\text{i}L=1.4$ match exactly with the results of Refs. 20, 21 and 7, respectively. Also, the variation of threshold pump intensity with $\alpha_\text{i}L$ is approximately linear which is consistent with the results of Ref. 22.

The above model accurately predicts the experimental results for weak idler absorption.$^7$ However, the model does not remain valid for the strong idler absorption, as reported in Ref. 10, due to large thermal load in the device.$^9$ Thus, due to thermal (lensing) effects,$^8$ the conversion efficiency decreases. X. Zhang et. al.$^{12}$ had also reported through both numerical simulations and experiments that the absorption of idler wave inside the nonlinear crystal can strongly effect the performance of OPO. Hence, the above model is not practical. Moore and Koch$^{13}$ had suggested a two-crystal OPO with four dichroic mirrors to enable dumping the idler wave that is generated in the first crystal. But, the use of several mirrors makes the device highly unstable and more lossy; and also the proposed ring cavity has some longer optical path, and hence the build-up time is larger as compared to the linear cavity.$^{14}$ To overcome these practical limitations, we propose a linear cavity twin-crystal SR-OPO with an idler absorber or a band stop filter placed in between the two crystals. Due to the use of a localized absorber between the two crystals, OPO will be free from thermal loading and thus the thermal effects does not affect the performance of OPO.

\section{Twin-crystal SR-OPO with an idler absorber}

Figure 3 shows a schematic of twin-crystal configuration of SR-OPO with an idler absorber placed in between the two crystals. This configuration is more compact and stable as compared to that of Ref. 13. To find the conversion efficiency for the signal we determine the evolution of the three fields inside the crystals and the absorber.
\begin{figure}[th]
\centering
\includegraphics[scale=0.4]{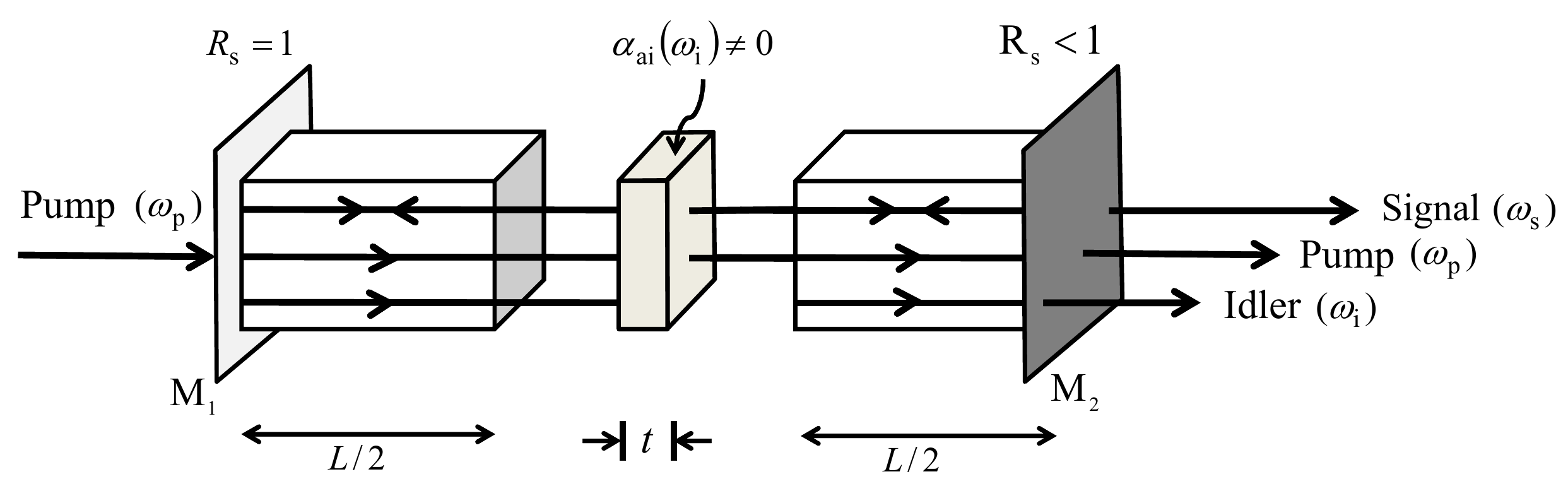}
\vspace*{8pt}
\caption{Schematic of twin-crystal configuration of SR-OPO with an idler absorber.}
\end{figure}

In the above analysis, the focusing effects of the three waves have been neglected. This is valid in practical cases. For example, for wavelengths in the range $\lambda =1-4\mu m$ and beam waist w$_\text{o}=1-3mm$, the angle of diffraction comes out to be few milliradians, and therefore one can assume the three beams to remain collimated and overlapping within the cavity. Further, Guha$^{23}$ had showed that under these approximations, the analytical expressions corresponding to focused beam interactions, reduced to plane wave results. 

Following a similar approach, as outlined in Sec. 2, we write coupled equations$^{18}$ for the case of perfect phase matching, and under the constant signal-wave approximation, for the first crystal:
\begin{equation}
\dfrac{dE_\text{1p}}{dz}=-i\kappa_\text{p}E_\text{1i}E_\text{1s}
,\label{this}
\end{equation}
\begin{equation}
\dfrac{dE_\text{1s}}{dz} \approx 0
,\label{this}
\end{equation}
\begin{equation}
\dfrac{dE_\text{1i}}{dz}=-i\kappa_\text{i}E_\text{1p}E^*_\text{1s}-\frac{\alpha_\text{i}}{2}E_\text{1i}
.\label{this}
\end{equation}
$E_\text{1m}(0 \leq z \leq L/2)$'s (m=p, s, i) represent the fields inside the first crystal. Using the boundary conditions:
\begin{equation}
E_\text{1p}(z=0)=E_\text{po} \quad \text{and} \quad E_\text{1i}(z=0)=0
,\label{this}
\end{equation}
we can solve Eqs. (3.1)-(3.3) to obtain the expressions for pump and idler amplitudes inside the first crystal. Since, there is no interaction among the three waves inside the absorber, propagation through the absorber would only introduce a phase factor in the three waves and a loss factor for the idler wave. For the second crystal also, we need to solve the same set of coupled equations as given by Eqs. (3.1)-(3.3). If $E_\text{am}(t)$'s (m= p, s, i) are the fields after passing through the absorber, then by using the boundary condition:
\begin{equation}
E_\text{2p}(z=0)=E_\text{ap}(t) \quad \text{and} \quad E_\text{2i}(z=0)=E_\text{ai}(t)
,\label{this}
\end{equation}
we get the following expression for the amplitude of the pump wave at the output of the OPO:
\begin{eqnarray}
E^{\text{out}}_\text{p}(z=0) &=& E_\text{po} \exp \left( i(\phi_\text{i}+\phi_\text{s}) \right) \nonumber \\
			   &  & \left[ \exp(i\Delta \phi)\cos^2\left( \frac{\beta L}{2}\right)-  \exp \left(-\frac{\alpha_\text{ai}t}{2}\right)\sin^2\left( \frac{\beta L}{2}\right) \right] ,\label{this}
\end{eqnarray}
where $\beta$ is the parametric gain of the OPO defined by Eq. (2.13) and $\Delta \phi=\phi_\text{p}-\phi_\text{s}-\phi_\text{i}$ is the relative inter-crystal phase shift of the three waves. $\phi_\text{m}$'s ( m = p, s, i) are the phase accumulated on the three waves in passing through the air gap and the absorber, and $\alpha_\text{ai}$ is the absorption coefficient for the idler wave inside the absorber. For $\Delta \phi=2q\pi$ , where $q$ is an integer, the conversion efficiency is maximum, and the following results correspond to this case. The significance of this term i.e. relative intercrystal phase shift will be discussed in Sec. 7. Above expression for the pump amplitude can be used to find the conversion efficiency for the signal using Eqs. (2.11) and (2.12) as shown in Sec. 2.

\begin{figure}[th]
\centering
\includegraphics[scale=0.4]{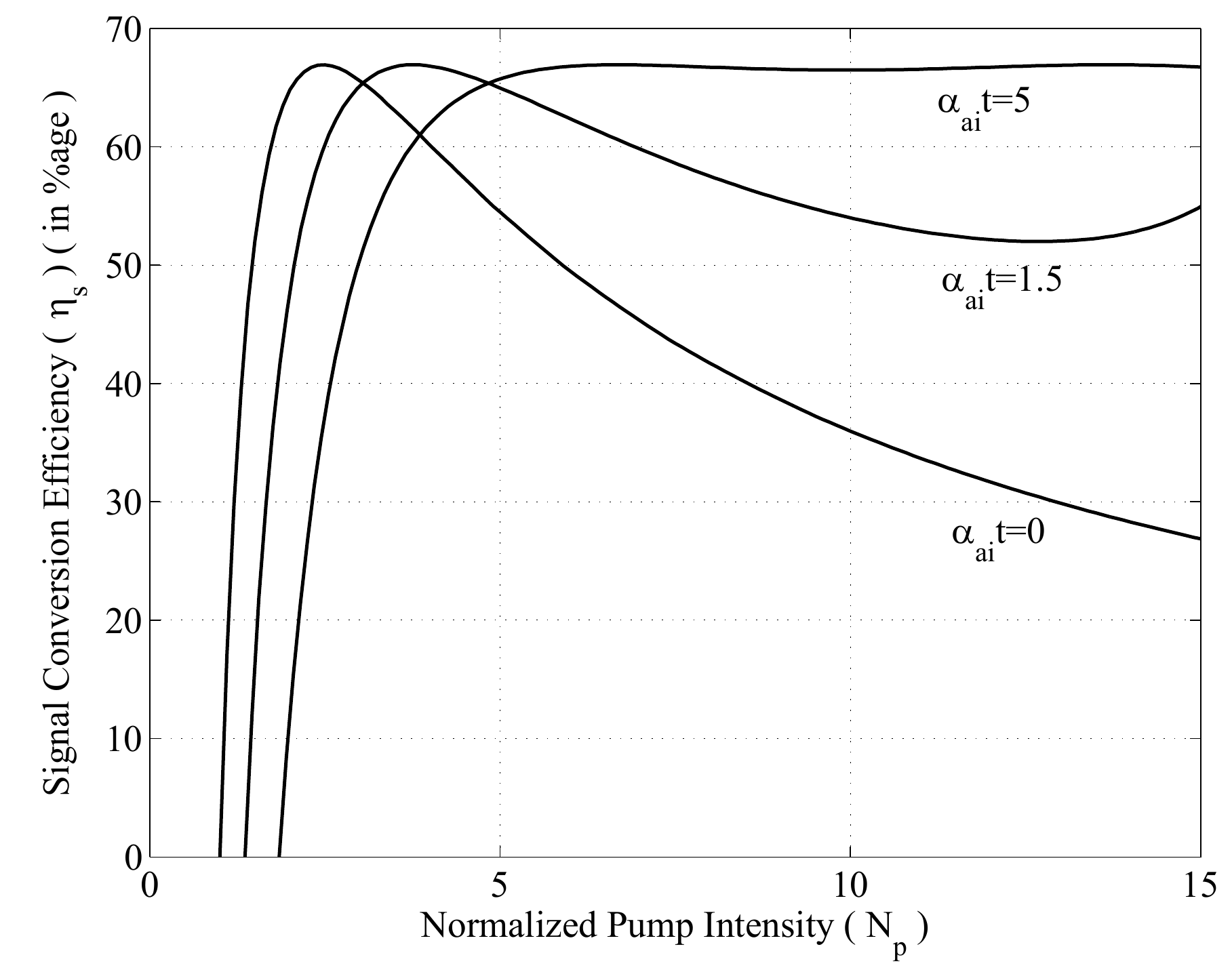}
\vspace*{8pt}
\caption{Variation of conversion efficiency for the signal with normalized pump intensity for different values of $\alpha_\text{ai}t$.}
\end{figure}

Figure 4 shows the variation of conversion efficiency for the signal with normalized pump intensity for different values of $\alpha_\text{ai}t$. It is clear that increase in absorption coefficient of the idler absorber leads to the suppression of back conversion without any thermal loading of the device.

Again we can determine the threshold pump intensity from the fact that near threshold $\beta L \rightarrow 0$.
\begin{equation}
N^{\text{th}}_\text{p}=\frac{I^{\text{th}}_\text{p}(\alpha_\text{ai}\neq 0)}{I^{\text{th}}_\text{p}(\alpha_\text{i}=0)}=\frac{2}{1+\exp{\left(-\frac{\alpha_\text{ai}t}{2} \right)}}
.\label{this}
\end{equation}

\begin{figure}
\centering
\includegraphics[scale=0.4]{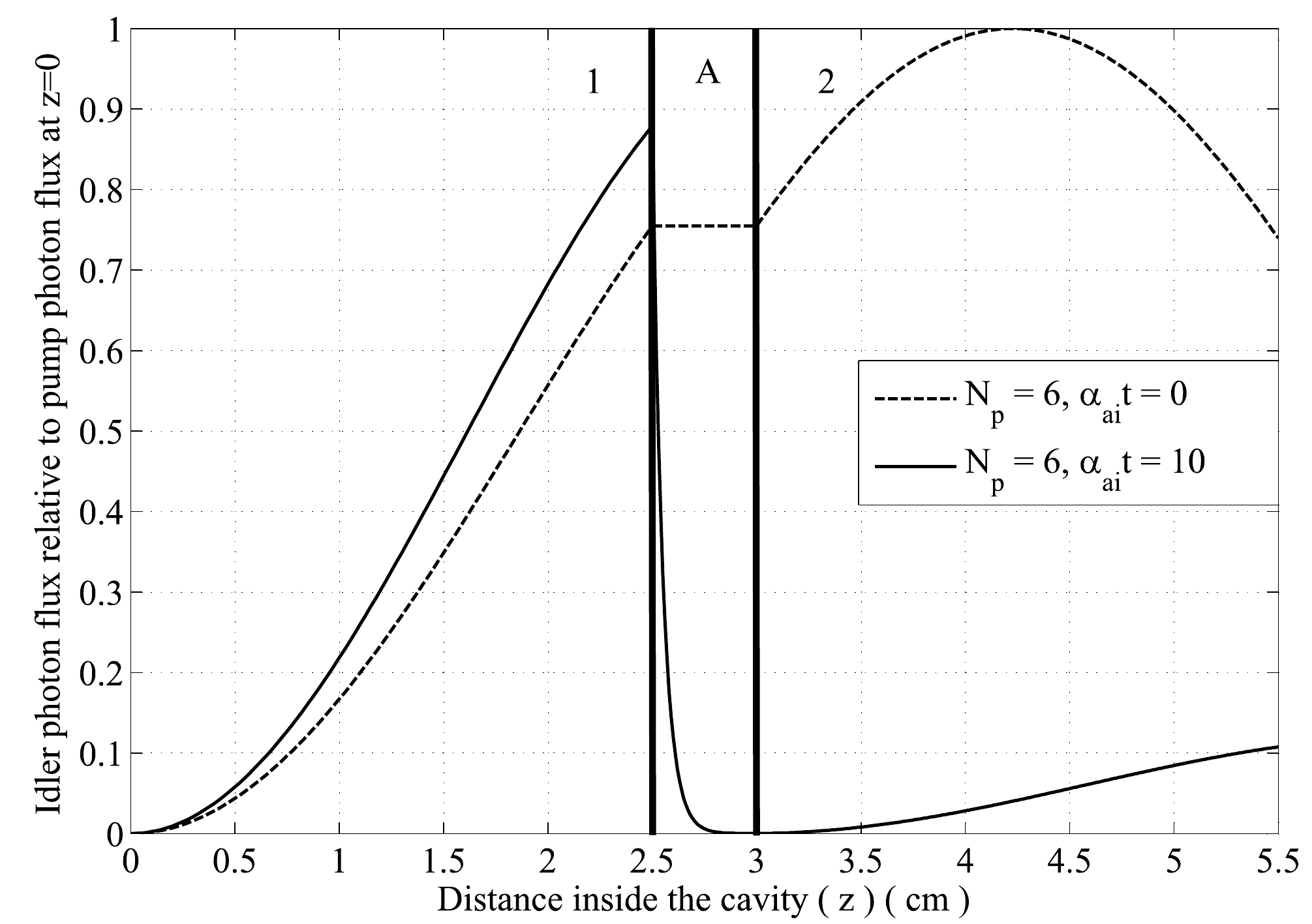}
\vspace*{20pt}
(a)
\includegraphics[scale=0.4]{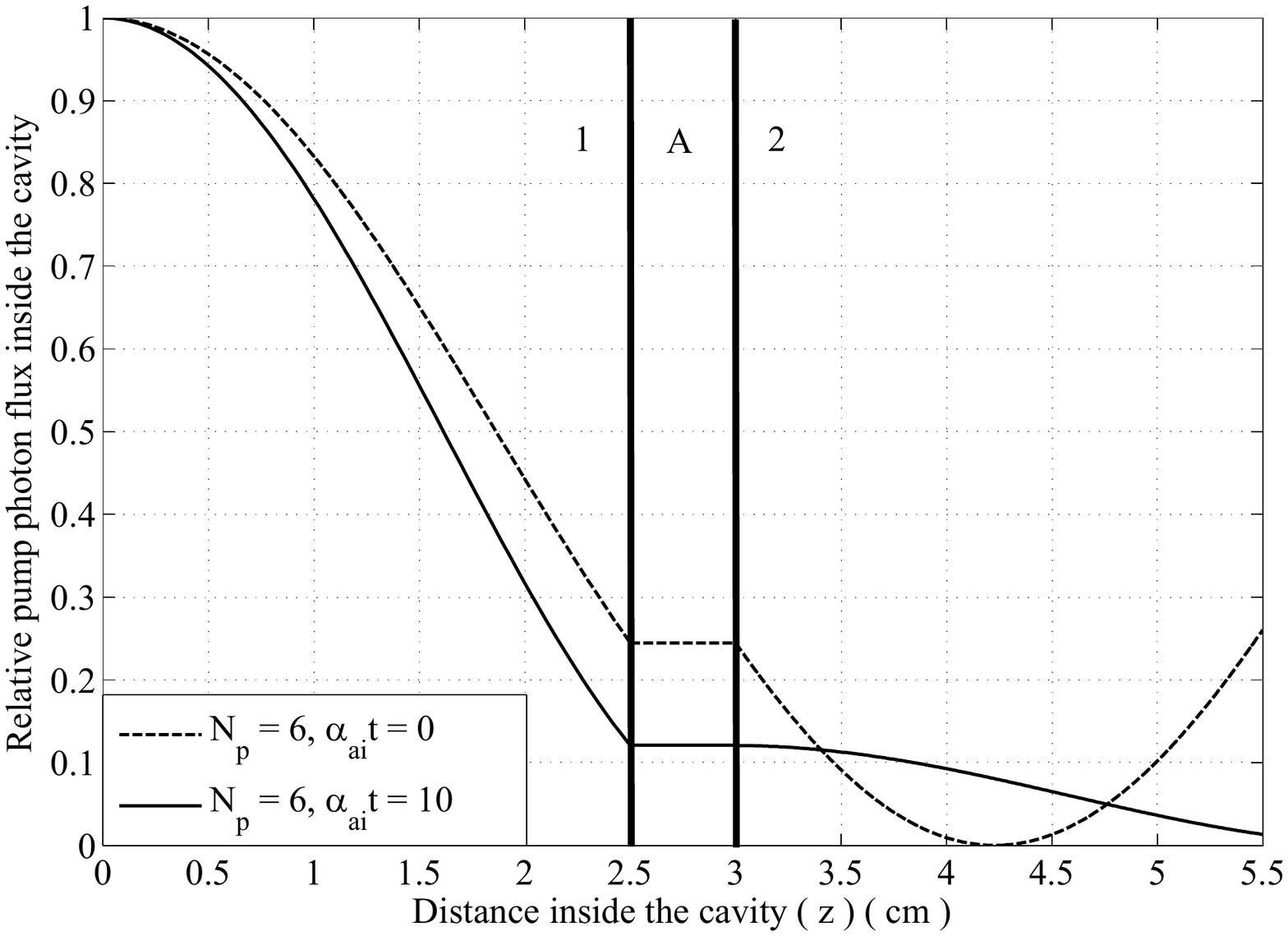}
(b)
\vspace*{10pt}
\caption{Variation of idler photon flux (a) and pump photon flux (b) relative to pump photon flux at $z=0$ inside the cavity for $\alpha_\text{ai}t=0,10$ with $N_\text{p}=6$.}
\end{figure}
Figures 5 shows the variation of idler and pump photon flux relative to the pump photon flux at $z=0$ inside the cavity for $\alpha_\text{ai}t=0$ and $10$ ; $z=0$ corresponds to the input end of the cavity and $z=L+A$ corresponds to the output end of the cavity, where $A$ is the intercrystal gap (including the thickness of the absorber) between the two crystals. Since the back conversion is effective beyond $N_\text{p}\approx2.5$, these variations have been plotted for $N_\text{p}=6$ (say). It is clear from the figures that after absorption of the idler photons inside the absorber (for $\alpha_\text{ai}t=10$), idler wave does not build up sufficiently to induce the back conversion in the second crystal and thus the pump wave keeps on depleting.

In the above we considered the use of an absorber in between the two crystals to dump the idler wave; however, a dichroic mirror can also be used in place of an idler absorber. Figure 6 shows the schematic of twin-crystal SR-OPO with a dichroic mirror M$_3$ sandwiched between the two crystals to dump the idler wave. The analysis will remain same, except $\exp(-\alpha_\text{ai}t)$ will be replaced by $1-R_\text{i}$ , where $R_\text{i}$ is the reflectivity of the dichroic mirror at the idler wavelength.
\begin{figure}[th]
\centering
\includegraphics[scale=0.4]{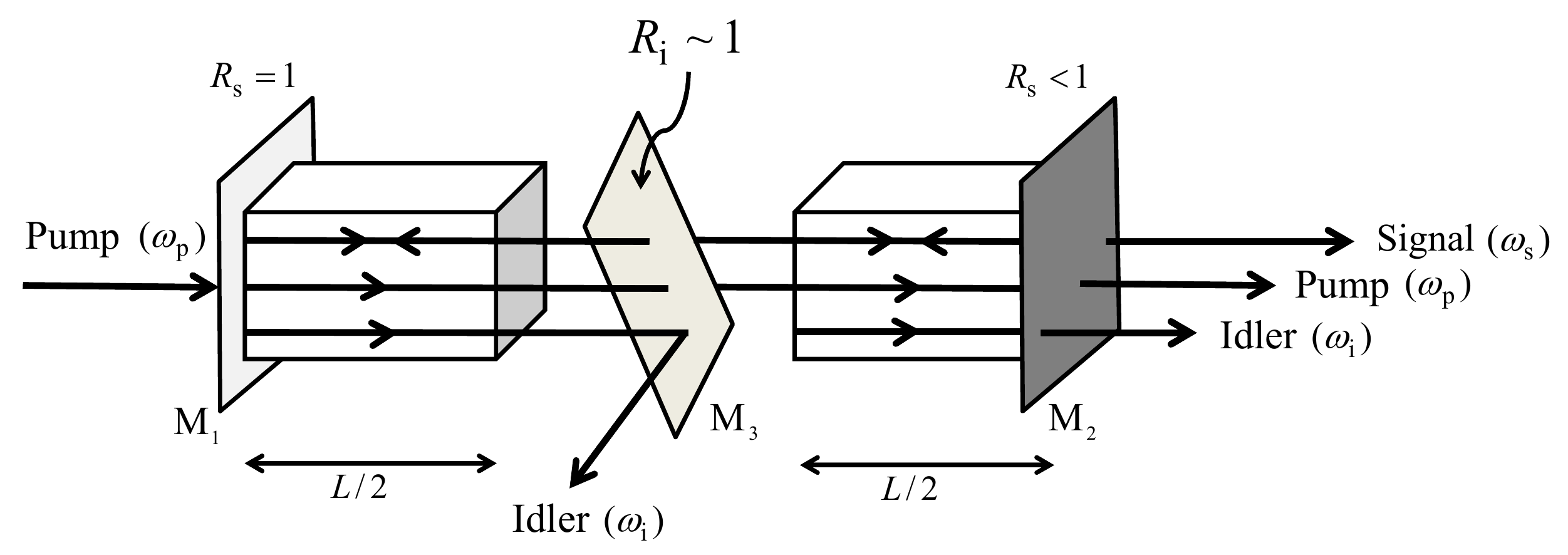}
\vspace*{8pt}
\caption{Schematic of twin-crystal configuration of SR-OPO with a dichroic mirror used to dump the idler wave generated inside the first crystal.}
\end{figure}
In practice, tunable OPOs are more interesting to generate idler wave (in the mid-IR) than the signal wave. For this, one can design an idler resonating OPO, and introduce a signal absorber located mid-way between the two crystals. Following the analysis presented above, one can solve the coupled wave equations to find the variation of idler conversion efficiency with the normalized pump intensity. The variation will be similar to Fig. 4, except the scaling of y-axis.
\section{Two-crystal SR-OPO with an idler absorber}
In the previous section we have taken the two crystals to be of equal length. In this section we will vary the position of the absorber relative to the crystal length and see its effect on the conversion efficiency for the signal. We follow the same procedure as in the previous section to solve the coupled differential equations. If $L_1$ and $L_2$ are the lengths of the first and the second crystal, respectively, with $L=L_1+L_2$, then the amplitude of the pump wave at the output of the second crystal is given by:
\begin{eqnarray}
E^\text{out}_\text{p}(z=0) &=& E_\text{po} \exp \left( i(\phi_\text{i}+\phi_\text{s}) \right) \nonumber \\
			   &  & \left[ \cos\left( \beta L_1\right) \exp(i\Delta \phi)\cos\left( \beta L_2\right)- \sin\left( \beta L_1\right) \exp \left(-\frac{\alpha_\text{ai}t}{2}\right)\sin\left( \beta L_2\right) \right]. \nonumber \\
\label{this}
\end{eqnarray}
\begin{figure}[th]
\centering
\includegraphics[scale=0.45]{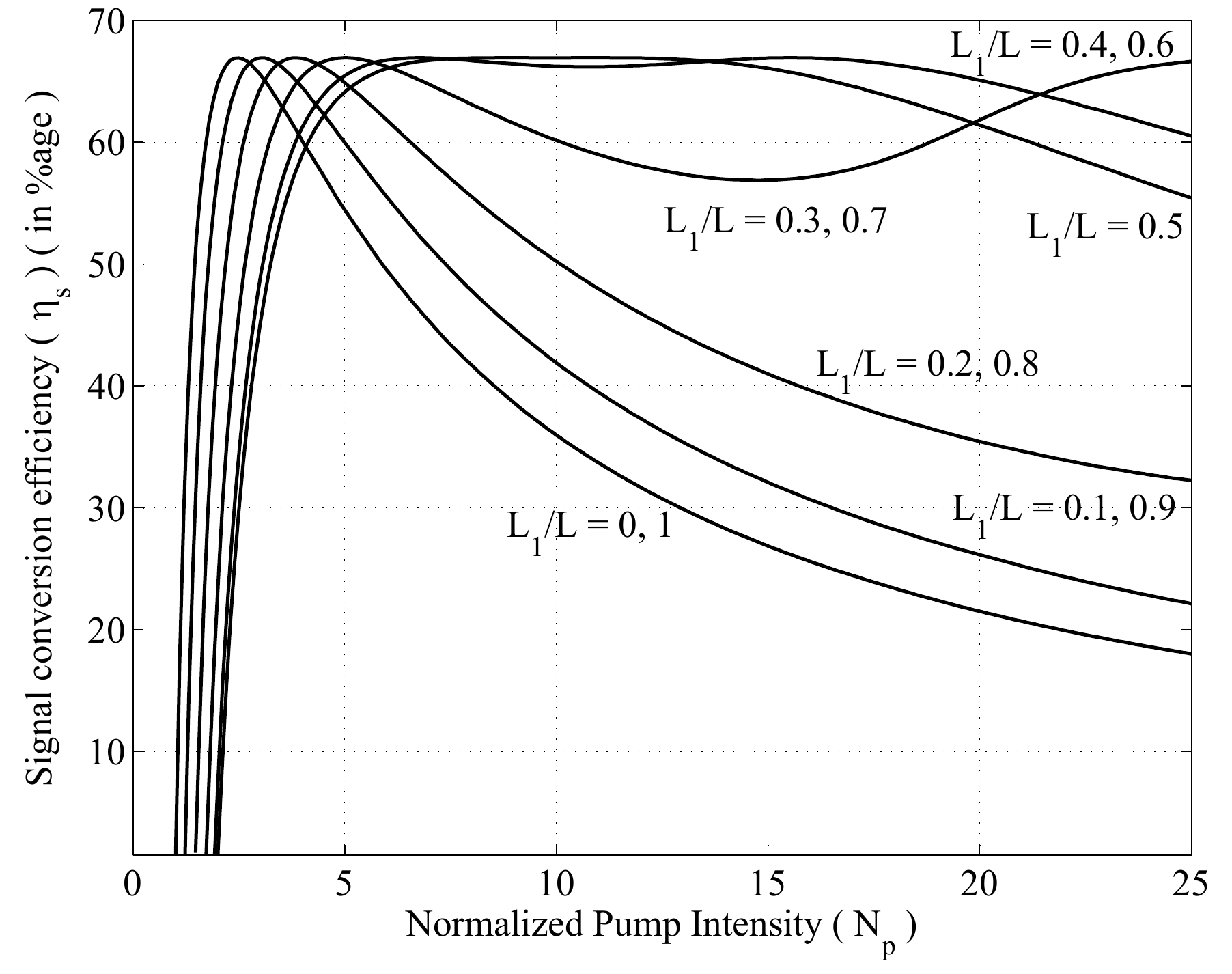}
\vspace*{20pt}
(a)
\includegraphics[scale=0.45]{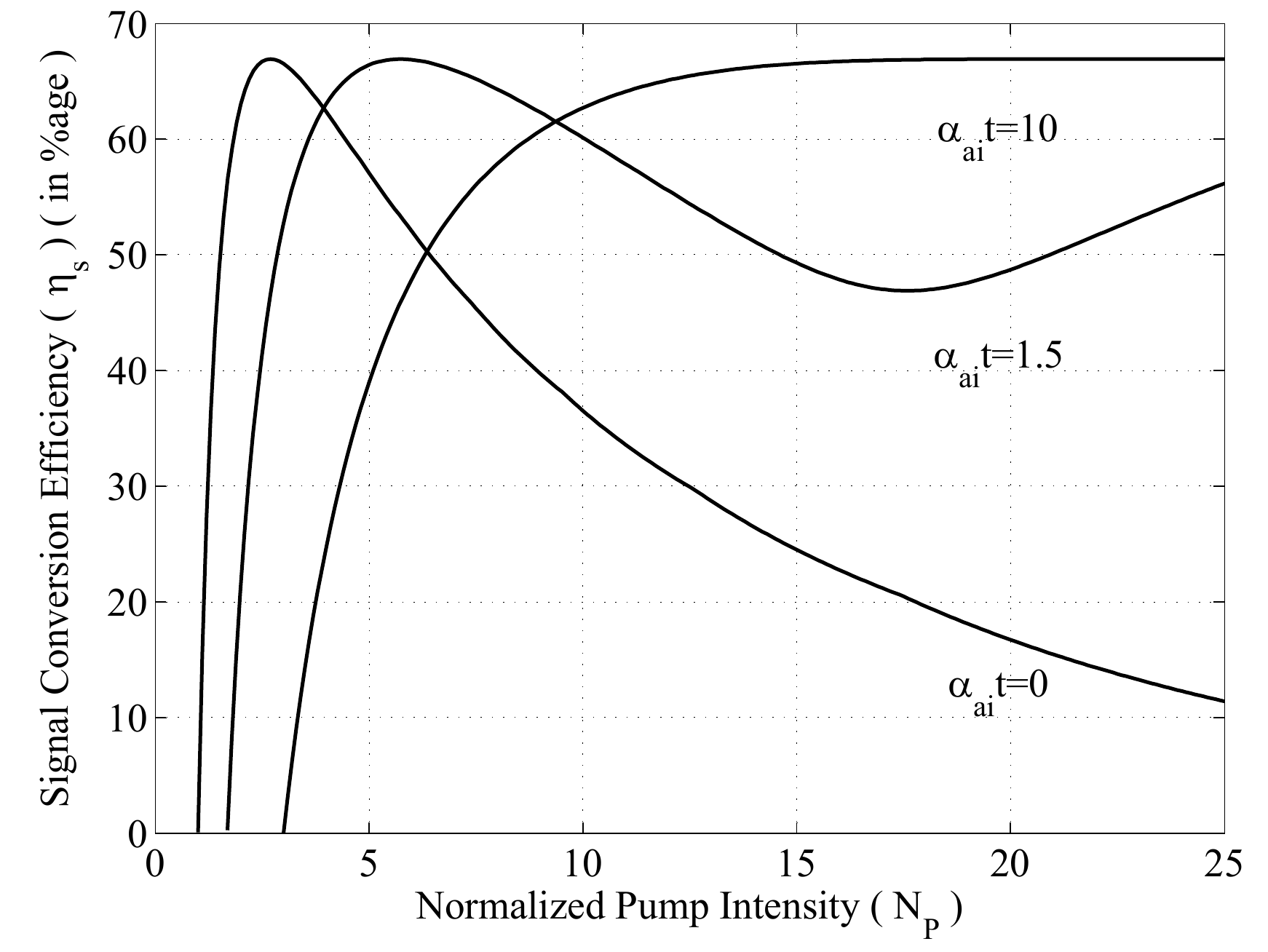}
\vspace*{10pt}
(b)
\caption{(a) Variation of conversion efficiency for the signal with normalized pump intensity for different relative positions of the absorber with $\alpha_{\text{ai}}t=10$, and (b) Variation of conversion efficiency for the signal with normalized pump intensity for three crystal SR-OPO with two idler absorbers at different values of $\alpha_\text{ai}t$.}
\end{figure}
Since near threshold $\beta L \rightarrow 0$ , normalized threshold pump intensity as a function of the position of the absorber is given by:
\begin{equation}
N^{\text{th}}_\text{p}=\frac{L^2}{L^2_1+L^2_2}
.\label{this}
\end{equation}
By considering $\Delta \phi=2q\pi$, the conversion efficiency for the signal can be determined from Eqs. (2.12), (2.13) and (4.1). Figure 7(a) shows the variation of the conversion efficiency for the signal for different positions of the absorber from which we can conclude that the back conversion is almost suppressed for the case of absorber placed nearly in mid-way between the two crystals. Therefore, it is useful to take the two crystals of equal length i.e., $L_1=L_2=L/2$.

\section{Three-crystal SR-OPO with an idler absorber}
For the twin-crystal configuration of SR-OPO with an idler absorber in between we have seen from Fig. 7(a) that for $N_\text{p}>15$ , conversion efficiency for the signal again starts dropping down due to the rapid growth of idler wave and back conversion to the pump wave inside the first crystal. At such high pump intensities, one can use three-crystal configuration of SR-OPO with strong idler absorbers placed at $L/3$ and $2L/3$ . In order to find the conversion efficiency and normalized threshold, we find the evolution of the three waves inside the cavity by following the same procedure outlined in Sec. 3.

Figure 7(b) shows the variation of conversion efficiency for the signal with normalized pump intensity for different values of $\alpha_\text{ai}t$ . It is clear from figure that three-crystal configuration is useful in suppressing the back conversion for $N_\text{p}>15$ . (cf. Fig. 6). From Fig. 4 and 7(b), one can conclude that the range of interest of $N_\text{p}$ will decide, whether one must use a twin-crystal SR-OPO with an idler absorber, or a three-crystal SR-OPO with two-idler absorbers.

\section{Optimum normalized pump intensity for a given value of $\alpha_\text{ai}t$}
In Fig. 4, one can see that as $\alpha_\text{ai}t$ increases, the pump intensity at which the conversion efficiency for the signal reaches maximum, also increases. Hence there exist an optimum normalized pump intensity $N^{\text{opt}}_\text{p}$ (at which we get maximum conversion efficiency for the signal) for a given $\alpha_\text{ai}t$ and the input pump intensities of interest are always less than the damage threshold of the nonlinear crystals.$^{24}$
\begin{figure}[th]
\centering
\includegraphics[scale=0.45]{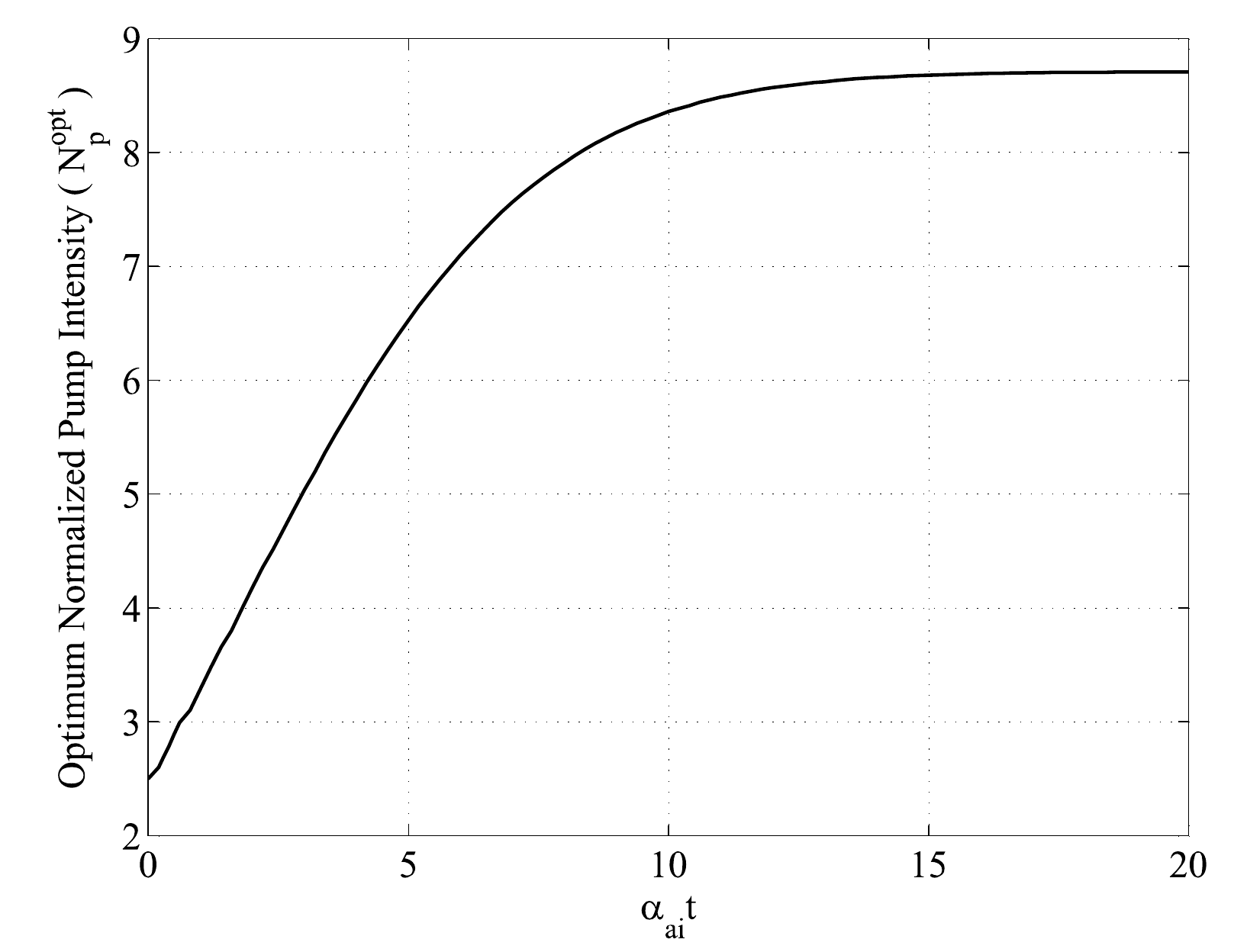}
\vspace*{8pt}
\caption{Variation of optimum normalized pump intensity with $\alpha_\text{ai}t$ ($n$ is an integer).}
\end{figure}
Figure 8 shows the variation of $N^{\text{opt}}_\text{p}$ with $\alpha_\text{ai}t$. As $\alpha_\text{ai}t$ increases, the $N^{\text{opt}}_\text{p}$ also increases; and for $\alpha_\text{ai}t>10$, the idler wave generated inside the first crystal is completely absorbed which leads to the saturation of $N^{\text{opt}}_\text{p}$.

\section{Relative inter-crystal phase shift}
In Sec. 3, 4 and 5 while calculating the conversion efficiency for the signal, we have assumed that the relative inter-crystal phase shift, $\Delta \phi=2q\pi$ where $q$ is an integer. In this section we discuss the effect of $\Delta \phi$ on the conversion efficiency for the signal.

Relative inter-crystal phase shift ($\Delta \phi$) is an important parameter that can be used to cancel the effect of phase mismatch in the three interacting waves.$^{15,16}$ For example, if the three waves are not perfectly phase matched in the first crystal and in the second crystal they are perfectly phase matched. So the three waves leave the first crystal with a definite phase relation between them. If the three waves enter in the second crystal with such a relative phase (other than $2q\pi$), then this may also reduce the interaction among the three waves inside the second crystal. However, if we use an inter-crystal spacer that introduces an additional relative phase among the three waves in such a way that the total relative phase (due to finite phase mismatch in the first crystal and due to the spacer) is $2q\pi$, and then there is maximum interaction among the three waves in the second crystal.

Relative inter-crystal phase shift can also be used to cancel the effect of opposite signs of effective nonlinear coefficients in the two crystals.$^{15,16}$ Change in sign of effective nonlinear coefficient implies an additional relative phase of $\pi$ among the three waves. If the two crystals have opposite signs of effective nonlinear coefficients, then by using an appropriate spacer we can introduces an additional relative inter-crystal phase shift of $\pi$ among the three waves. So the total relative phase of three waves again becomes $2q\pi$ that leads to maximum interaction.

If at least one of the three interacting waves is absent at the input of second crystal then there is no effect of relative inter-crystal phase shift on the conversion efficiency for the signal.$^{13}$ Thus, if we introduce an idler absorber as the spacer with $\alpha_\text{ai}t>10$ , then the whole idler wave gets absorbed inside the absorber and at the input of second crystal there is no idler wave. So, the variation of the conversion efficiency for the signal is insensitive to the variation in the relative inter-crystal phase shift.
\begin{figure}[th]
\centering
\includegraphics[scale=0.45]{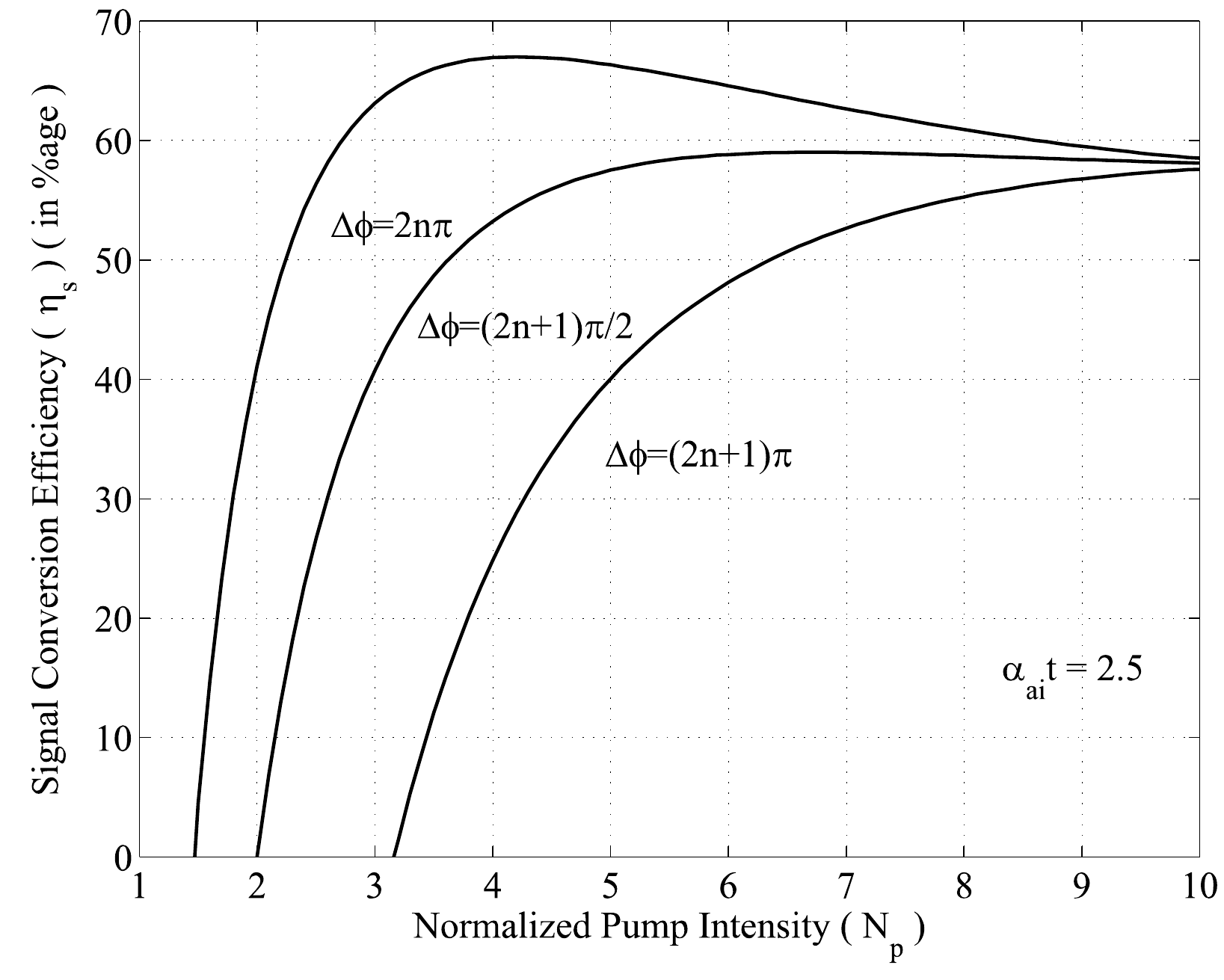}
\vspace*{8pt}
\caption{Variation of conversion efficiency for the signal with normalized pump intensity for different values of relative inter-crystal phase shift with $\alpha_\text{ai}t=2.5$ ($n$ is an integer).}
\end{figure}
Figure 9 shows the variation of conversion efficiency for the signal with normalized pump intensity for different values of relative inter-crystal phase shift with $\alpha_\text{ai}t=2.5$, which reflects that for $\Delta \phi=2q\pi$ conversion efficiency for the signal is maximum.
\begin{figure}[th]
\centering
\includegraphics[scale=0.4]{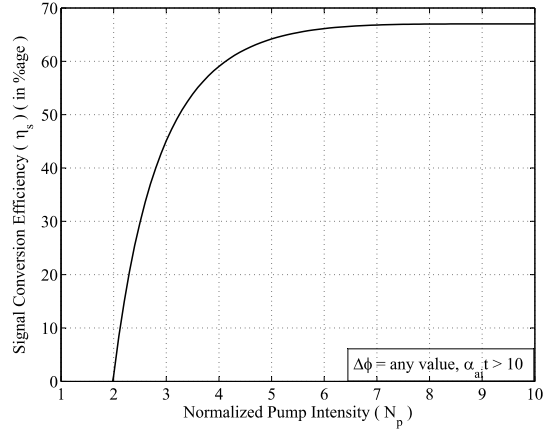}
\vspace*{8pt}
\caption{Variation of conversion efficiency for the signal with normalized pump intensity for any relative inter-crystal phase shift with $\alpha_\text{ai}t>10$.}
\end{figure}

Figure 10 shows the variation of conversion efficiency for the signal with normalized pump intensity for all values of the relative inter-crystal phase shift with $\alpha_\text{ai}t>10$. From which we concludes that the variation of conversion efficiency for the signal is insensitive to the variation of relative inter-crystal phase shift for all $\alpha_\text{ai}t>10$.
\section{Steady state analysis of twin-crystal SR-OPO without constant signal-wave approximation}
The above discussion for twin-crystal SR-OPO with an idler absorber was based on the constant signal-wave approximation, where the intensity of the signal wave is assumed to be constant inside the cavity. In this section, we present the steady state analysis for twin-crystal SR-OPO with an idler absorber, and have shown that the constant signal-wave approximation holds good if the reflectivity of the output coupler is high $R_\text{s}>0.7$. When the reflectivity of the output coupler is not too high, the variations of the three fields can be described by elliptic functions. Following Bey and Tang$^{17}$, we have obtained the numerical results for the conversion efficiency for the signal with input pump intensity for different transmission coefficients of the output coupler. In this analysis, we have used an idler absorber of infinitesimal thickness and very high idler absorption coefficient, between the two crystals each of length $L/2$. The output coupler is placed at the end of the second crystal.

Assuming perfect phase matching of the three waves inside the two crystals, coupled differential equations governing the nonlinear interaction of three waves in the first crystal are given by:
\begin{equation}
\dfrac{dE_\text{1p}}{dz}=-i\kappa_\text{p}E_\text{1i}E_\text{1s}
,\label{this}
\end{equation}
\begin{equation}
\dfrac{dE_\text{1s}}{dz}=-i\kappa_\text{s}E_\text{1p}E^*_\text{1i}
,\label{this}
\end{equation}
\begin{equation}
\dfrac{dE_\text{1i}}{dz}=-i\kappa_\text{i}E_\text{1p}E^*_\text{1s}
,\label{this}
\end{equation}
where $E_\text{1m}$'s (m=p, s, i) are amplitudes of the three waves inside the first crystal. Writing the amplitudes in the
following form:
\begin{equation}
E_\text{1m}(z)=\xi_\text{1m}(z)\exp{(i\phi_\text{1m}(z))}
,\label{this}
\end{equation}
and, defining the normalized amplitude of the three waves as:
\begin{equation}
u_\text{1m}=\left(\frac{n_\text{m}\epsilon_\text{o}c}{2\omega_\text{m}I}\right)^{1/2}\xi_\text{1m}
,\label{this}
\end{equation}
where $I=\frac{1}{2}c\epsilon_\text{o}(n_\text{i}\xi^2_\text{1i}+n_\text{s}\xi^2_\text{1s}+n_\text{p}\xi^2_\text{1p})$ is the total intensity of the three waves which is a conserved quantity; $u^2_\text{1m}$'s are proportional to the photon flux of the three waves in the first
crystal. The normalized length ($\zeta_1$) is defined as:
\begin{equation}
\zeta_1=\frac{d_{\text{eff}}}{c}(n_\text{i}\xi^2_\text{1i}+n_\text{s}\xi^2_\text{1s}+n_\text{p}\xi^2_\text{1p})^{\frac{1}{2}}\left( \frac{\omega_\text{i}\omega_\text{s}\omega_\text{p}}{n_\text{i}n_\text{s}n_\text{p}} \right)^{\frac{1}{2}}z
,\label{this}
\end{equation}
where $0\leq z\leq L/2$ and $0\leq \zeta_1 \leq l/2$
The real and imaginary parts of Eqs. (8.1)-(8.3) can now be written in the form:
\begin{equation}
\dfrac{du_\text{1p}}{d\zeta_1}=-u_\text{1i}u_\text{1s}\sin\theta_1
,\label{this}
\end{equation}
\begin{equation}
\dfrac{du_\text{1s}}{d\zeta_1}=u_\text{1p}u^*_\text{1i}\sin\theta_1
,\label{this}
\end{equation}
\begin{equation}
\dfrac{du_\text{1i}}{d\zeta_1}=u_\text{1p}u^*_\text{1s}\sin\theta_1
,\label{this}
\end{equation}
and
\begin{equation}
\dfrac{d\theta_1}{d\zeta_1}=\left( \frac{\kappa_\text{i} \xi_\text{1p} \xi_\text{1s}}{\xi_\text{1i}}+\frac{\kappa_\text{s} \xi_\text{1p} \xi_\text{1i}}{\xi_\text{1s}}+\frac{\kappa_\text{p} \xi_\text{1i} \xi_\text{1s}}{\xi_\text{1p}} \right) \cos\theta_1
,\label{this}
\end{equation}
respectively,
where $\theta_1=\phi_\text{1p}-\phi_\text{1s}-\phi_\text{1i}$.
With the boundary condition:
\begin{equation}
u_\text{1i}(z=0)=0
.\label{this}
\end{equation}
Manley-Rowe relations$^{18}$ for the first crystal can be written as:
\begin{equation}
u^2_\text{1s}(0)+u^2_\text{1p}(0)=u^2_\text{1s}(\zeta_1)+u^2_\text{1p}(\zeta_1)
,\label{this}
\end{equation}
\begin{equation}
u^2_\text{1p}(0)=u^2_\text{1p}(\zeta_1)+u^2_\text{1i}(\zeta_1)
,\label{this}
\end{equation}
\begin{equation}
u^2_\text{1s}(0)=u^2_\text{1s}(\zeta_1)+u^2_\text{1i}(\zeta_1)
.\label{this}
\end{equation}
Signal photon flux in the first crystal in terms of Jacobian elliptic functions$^{25,26}$ is given by the following expression:
\begin{eqnarray}
u^2_\text{1s}(0 \leq \zeta_1 \leq l/2)&=&u^2_\text{1s}(0)+u^2_\text{1p}(0) \nonumber \\
								  & &\left[ 1-\text{sn}^2 \left[ (u^2_\text{1s}(0)+u^2_\text{1p}(0))^{\frac{1}{2}}|\zeta_1-\zeta_\text{o}| , \left(\frac{u^2_\text{1p}(0)}{u^2_\text{1s}(0)+u^2_\text{1p}(0)}\right)^{\frac{1}{2}} \right] \right] \nonumber \\
\end{eqnarray}
Signal photon flux at the output of the first crystal is:
\begin{equation}
u^2_\text{1s}\left(\frac{l}{2}\right)=u^2_\text{1s}(0)+u^2_\text{1p}(0)\left[1-A\right]
,\label{this}
\end{equation}
where
\begin{equation}
A=\text{sn}^2 \left[ (u^2_\text{1s}(0)+u^2_\text{1p}(0))^{\frac{1}{2}}\left|\frac{l}{2}-\zeta_\text{o}\right| , \left(\frac{u^2_\text{1p}(0)}{u^2_\text{1s}(0)+u^2_\text{1p}(0)}\right)^{\frac{1}{2}} \right]
.\label{this}
\end{equation}
The parameter $\zeta_\text{o}$ is determined from the fact that at $\zeta_1=0$, Eq. (8.15) is also valid:
\begin{equation}
1=\text{sn}^2 \left[ (u^2_\text{1s}(0)+u^2_\text{1p}(0))^{\frac{1}{2}}\zeta_\text{o} , \left(\frac{u^2_\text{1p}(0)}{u^2_\text{1s}(0)+u^2_\text{1p}(0)}\right)^{\frac{1}{2}} \right]
.\label{this}
\end{equation}
Using Eqs. (8.12)-(8.15) we obtain the expression for the photon flux of the three waves. To find the evolution of photon flux of the three waves inside the second crystal, we follow the same approach, and solve the coupled differential equations. Due to the depletion of pump wave inside the first crystal and the presence of an idler absorber (of infinitesimal thickness) just after the first crystal that absorbs the all idler photons, boundary conditions become:
\begin{equation}
u^2_\text{2p}(0)l'^2=u^2_\text{1p}\left(\frac{l}{2}\right)l^2
,\label{this}
\end{equation}
\begin{equation}
u^2_\text{2s}(0)l'^2=u^2_\text{1s}\left(\frac{l}{2}\right)l^2
,\label{this}
\end{equation}
\begin{equation}
u^2_\text{2i}(0)=0
,\label{this}
\end{equation}
where $u^2_\text{2m}(0)$'s are proportional to the photon flux of the three waves at the input of second crystal and $l'$ is defined through the normalized length ($\zeta_2$) for the second crystal:
\begin{equation}
\zeta_2=\frac{d_{\text{eff}}}{c}(n_\text{i}\xi^2_\text{2i}+n_\text{s}\xi^2_\text{2s}+n_\text{p}\xi^2_\text{2p})^{\frac{1}{2}}\left( \frac{\omega_\text{i}\omega_\text{s}\omega_\text{p}}{n_\text{i}n_\text{s}n_\text{p}} \right)^{\frac{1}{2}}z
,\label{this}
\end{equation}
and $0\leq z\leq L/2$, gives $0\leq \zeta_2 \leq l'/2$. Using the Manley-Rowe relations and by solving the coupled differential equations for the second crystal, we obtain the following expression of the signal photon flux in the second crystal:
\begin{eqnarray}
u^2_\text{2s}(0 \leq \zeta_2 \leq l'/2)&=&u^2_\text{2s}(0)+u^2_\text{2p}(0) \nonumber \\
								  & &\left[ 1-\text{sn}^2 \left[ (u^2_\text{2s}(0)+u^2_\text{2p}(0))^{\frac{1}{2}}|\zeta_2-\zeta'_\text{o}| , \left(\frac{u^2_\text{2p}(0)}{u^2_\text{2s}(0)+u^2_\text{2p}(0)}\right)^{\frac{1}{2}} \right] \right]. \nonumber \\
\end{eqnarray}
Signal photon flux at the output of the second crystal is given by:
\begin{eqnarray}
u^2_\text{2s}\left(\frac{l'^2}{2l}\right)^2&=&u^2_\text{1s}(0)+u^2_\text{1p}(0) \nonumber \\
								  & &\left[ 1-A\,\text{sn}^2 \left[ (u^2_\text{1s}(0)+u^2_\text{1p}(0))^{\frac{1}{2}}\left|\frac{l}{2}-\zeta_\text{o}''\right| , \left(\frac{u^2_\text{1p}(0)A}{u^2_\text{1s}(0)+u^2_\text{1p}(0)}\right)^{\frac{1}{2}} \right] \right], \nonumber \\
\end{eqnarray}
where 
\begin{eqnarray}
\zeta'_\text{o}=\frac{l}{l'}\zeta_\text{o}'' , \nonumber
\end{eqnarray}
where $\zeta_\text{o}''$ is the parameter, determined from the fact that the Eq. (8.23) must be valid at $\zeta_2=0$ also, that gives:
\begin{equation}
1=\text{sn}^2 \left[ (u^2_\text{1s}(0)+u^2_\text{1p}(0))^{\frac{1}{2}}\zeta_\text{o}'' , \left(\frac{u^2_\text{1p}(0)A}{u^2_\text{1s}(0)+u^2_\text{1p}(0)}\right)^{\frac{1}{2}} \right]
.\label{this}
\end{equation}
Since there is no parametric gain for the propagation of signal wave in the backward direction, for steady state, the signal photon flux at the input of the first crystal and at the output of the second crystal are related through:
\begin{equation}
u^2_\text{1s}(0)l^2=R_\text{s}u^2_\text{2s}\left(\frac{l'}{2}\right)l'^2
.\label{this}
\end{equation}
Thus, Eq. (8.24) becomes:
\begin{eqnarray}
u^2_\text{1s}(0)&=&\left(\frac{R_\text{s}}{1-R_\text{s}}\right)u^2_\text{1p}(0) \nonumber \\
		   & &\left[ 1-A\,\text{sn}^2 \left[ (u^2_\text{1s}(0)+u^2_\text{1p}(0))^{\frac{1}{2}}\left|\frac{l}{2}-\zeta_\text{o}''\right| , \left(\frac{u^2_\text{1p}(0)A}{u^2_\text{1s}(0)+u^2_\text{1p}(0)}\right)^{\frac{1}{2}} \right] \right]. \nonumber \\
\end{eqnarray}
Eqs. (8.18), (8.25) and (8.27) contain the whole information which is required to determine the general steady state behavior of the twin-crystal configuration of SR-OPO with an idler absorber. These three equations can be solved numerically to obtain $u^2_\text{1s}(0)/u^2_\text{1p}(0)$, for a given $u_\text{1p}(0)l$, which is related to the input pump intensity by the following expression:
\begin{equation}
u_\text{1p}(0)l=\sqrt{\frac{2\omega_\text{i}\omega_\text{s}}{\epsilon_\text{o}c^3n_\text{p}n_\text{s}n_\text{i}}I_\text{p}(0)}d_{\text{eff}}L
.\label{this}
\end{equation}
For the variation of threshold pump intensity with the transmission coefficient of the output coupler, one can assume the pump to be undepleted and can solve the coupled differential equations to obtain:
\begin{equation}
(1-T)\cosh^4\left(  \left[ u_\text{1p}(0)l/2  \right]_{\text{threshold}} \right)=1
.\label{this}
\end{equation}
Now the conversion efficiency for the signal is given by:
\begin{equation}
\eta_\text{s}=\left(\frac{1-R_\text{s}}{R_\text{s}}\right)\left(\frac{u^2_\text{1s}(0)}{u^2_\text{1p}(0)}\right)
.\label{this}
\end{equation}
\begin{figure}[th]
\centering
\includegraphics[scale=0.4]{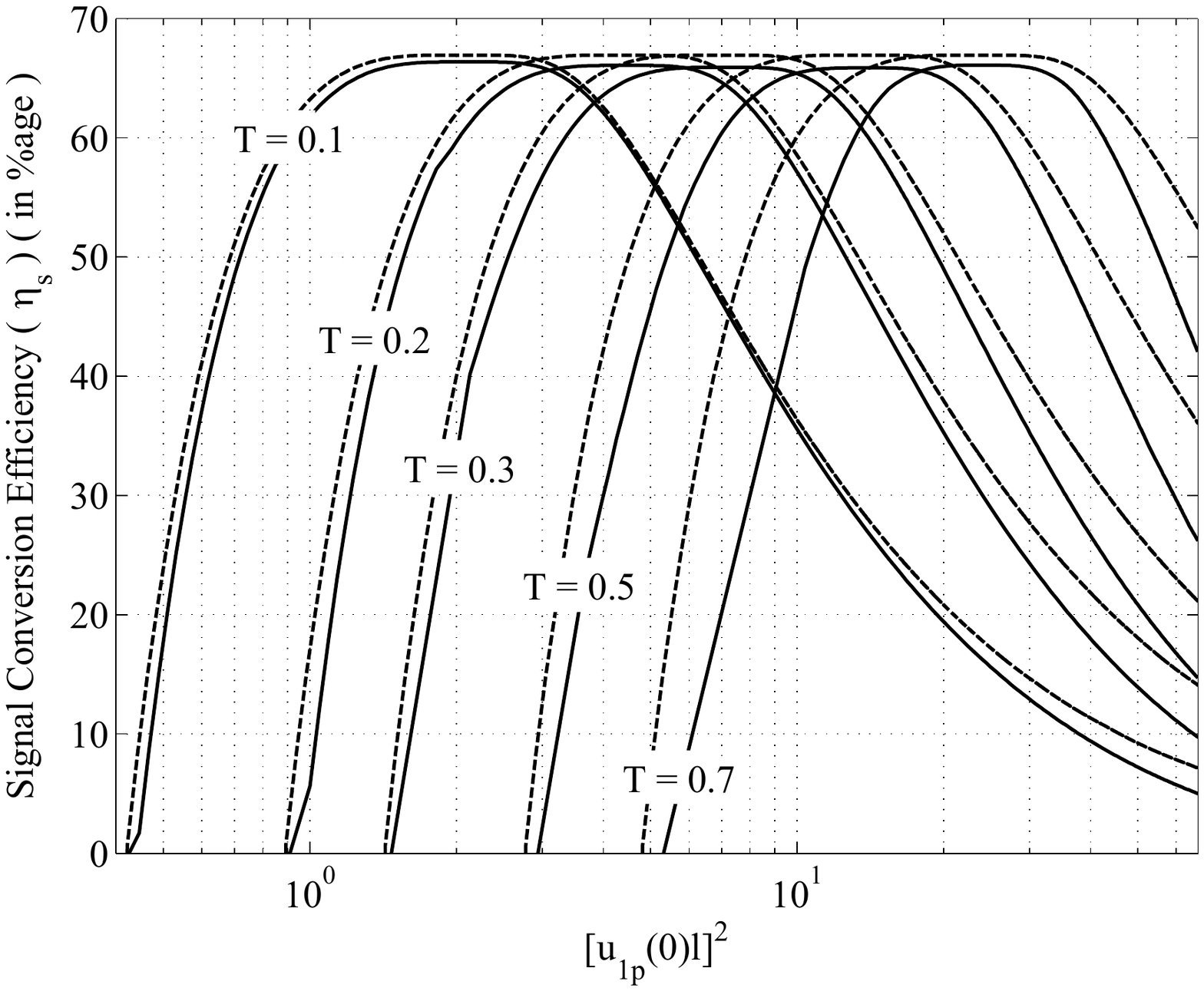}
\vspace*{8pt}
\caption{Variation of conversion efficiency for the signal with $u^2_\text{1p}l^2$, which is proportional to the input pump intensity, for different transmission coefficients (T) of the output coupler. The dashed curves represent the variations obtained with constant signal-wave approximation and the solid lines are the variations for the present analysis.}
\end{figure}
If $R_\text{s}$ is large, the number of signal photons inside the cavity are much greater than that of the pump photons. In this limit, our result for the pump amplitude at the output of OPO coincides with that obtained in Sec. 2. Figure 11 shows the variation of conversion efficiency for the signal with $u^2_\text{1p}(0)l^2$ for different values of transmission coefficients (T) of the output coupler. As can be seen, the threshold increases with increasing T , and one can also estimate the optimum output coupling for a given pump intensity. The dashed curves are obtained under the constant signal-wave approximation, while the solid curves correspond to the present analysis. It is clear from the figure that the constant signal-wave approximation is valid even for twin-crystal configuration of SR-OPO with an idler absorber, if the reflectivity of the output coupler is high (i.e., for $R_\text{s}$ approximately $0.7$ or more).

\section{Conclusions}
We have presented the analysis of twin-crystal SR-OPO with an idler absorber located in between the two crystals. We have presented results for the conversion efficiency for the signal as a function of normalized pump intensity for different values of the absorption coefficient. We conclude that the introduction of an idler absorber between two crystals of equal lengths, leads to significant increase in the conversion efficiency for the signal because of the prevention of back
conversion. At higher pump intensities $(N_\text{p}>15)$, one can use three-crystal SR-OPO with two absorbers to enhance the conversion efficiency. By operating the OPO at an optimum pump intensity, for a given $\alpha_\text{ai}t$, one can obtain maximum conversion efficiency. To make the conversion efficiency independent of the relative inter-crystal phase shift $(\Delta \phi)$, we can use a strong idler absorber with $\alpha_\text{ai}t>10$; otherwise, the conversion efficiency will be maximum only when $\Delta \phi$ is an integral multiple of $2\pi$. The steady state analysis of twin-crystal SR-OPO with an idler absorber also shows that the often used constant signal-wave approximation is valid if reflectivity of the output coupler is high. An idler resonating SR-OPO, with a signal absorber, would lead to relatively large idler conversion efficiency, and hence large idler power output. This should be very useful for the generation of high power mid-IR radiation.

\section*{Acknowledgments}
Authors thank Dr. Sebabrata Mukherjee and Vikram Kamaljith for the helpful discussion.

\end{document}